\documentclass[aoas,MSNbibl,nameyear,dvips]{arximspdf}
\usepackage{dcolumn}
\usepackage{graphicx}
\usepackage{url,breakurl}

\doi{10.1214/14-AOAS772} 
\volume{8}
\issue{4}
\pubyear{2014}
\firstpage{2319}
\lastpage{2335}
\docsubty{FLA}

\makeatletter
\mathchardef\varLambda="0103
\newcolumntype{d}[1]{D{.}{.}{#1}}
\def\cal{\mathcal}
\newcommand{\bbeta}{\bolds{\beta}}
\newcommand{\btheta}{\bolds{\theta}}
\newcommand{\bSigma}{\bolds{\Sigma}}
\newcommand{\Y}{\mathbf{Y}}
\makeatother

\begin{document}
\begin{frontmatter}

\title{A permutational-splitting sample procedure to quantify expert
opinion on clusters of chemical compounds using high-dimensional
data\thanksref{T1}}
\runtitle{A permutational-splitting sample procedure}
\thankstext{T1}{Supported in part by the IAP research network \#P7/06
of the Belgian Government (Belgian Science Policy).}

\begin{aug}
\author[A]{\fnms{Elasma} \snm{Milanzi}\corref{}\thanksref{m1}\ead[label=e1]{elasma.milanzi@uhasselt.be}},
\author[B]{\fnms{Ariel} \snm{Alonso}\thanksref{m2}\ead[label=e2]{ariel.alonso@maastrichtuniversity.nl}},
\author[C]{\fnms{Christophe} \snm{Buyck}\thanksref{m3}\ead[label=e3]{cbuyck@its.jnj.com}},
\author[D]{\fnms{Geert} \snm{Molenberghs}\thanksref{m1,m4}\ead[label=e4]{geert.molenberghs@uhasselt.be}\ead[label=e7]{geert.molenberghs@med.kuleuven.be}}
\and
\author[C]{\fnms{Luc} \snm{Bijnens}\thanksref{m3}\ead[label=e5]{lbijnens@its.jnj.com}}
\runauthor{E. Milanzi et al.}
\affiliation{Hasselt University\thanksmark{m1}, Maastricht
University\thanksmark{m2},
Janssen Pharmaceuticals\thanksmark{m3} and~University~of~Leuven\thanksmark{m4}}
\address[A]{E. Milanzi\\
I-BioStat\\
Hasselt University\\
Martelarenlaan 42\\
3500 Hasselt\\
Belgium\\
\printead{e1}}
\address[B]{A. Alonso\\
Department of Methodology \& Statistics\\
Maastricht University\\
P.O. Box 616\\
6200 MD Maastricht\\
Netherlands\\
\printead{e2}}
\address[C]{C. Buyck\\
L. Bijnens\\
Janssen Pharmaceuticals\\
Turnhoutseweg 30 \\
2340 Beerse\\
Belgium\\
\printead{e3}\\
\phantom{E-mail:\ }\printead*{e5}}
\address[D]{G. Molenberghs\\
I-BioStat\\
Hasselt University\\
Martelarenlaan 42\\
3500 Hasselt\\
Belgium\\
and\\
I-BioStat\\
University of Leuven\\
Kapucijnenvoer 35, Blok D, bus 7001\hspace*{21pt}\\
3000 Leuven\\
Belgium\\
\printead{e4}\\
\phantom{E-mail:\ }\printead*{e7}}
\end{aug}

\received{\smonth{9} \syear{2012}}
\revised{\smonth{5} \syear{2014}}

%
\begin{abstract}
Expert opinion plays an important role when selecting promising
clusters of chemical compounds in the drug discovery process. We
propose a method to quantify these qualitative assessments using
hierarchical models. However, with the most commonly available
computing resources, the high dimensionality of the vectors of fixed
effects and correlated responses renders maximum likelihood unfeasible
in this scenario. We devise a reliable procedure to tackle this problem
and show, using theoretical arguments and simulations, that the new
methodology compares favorably with maximum likelihood, when the latter
option is available. The approach was motivated by a case study, which
we present and analyze.
\end{abstract}


\end{frontmatter}

\section{Introduction}\label{sec1}
\subsection{Motivating case study}\label{sec:case}
Janssen Pharmaceutica carried out a project to assess the potential of
$22\mbox{,}015$ clusters of chemical compounds to identify those that
warranted further screening. In total, $147$ experts took part in the
study. For the analysis, their assessments were coded as $1$ if the
expert recommended the cluster for inclusion in the sponsor's database
and $0$ otherwise.

The experts used the desk-top application Third Dimension Explorer
(3DX) and had no contact with one another during the evaluation
sessions [\citet{abcd07}].
In a typical session, an expert evaluated a subset of clusters selected
at random from the entire set of $22\mbox{,}015$. Each cluster was presented
with additional information that included its size, the structure of
some of its distinctive members such as the compound with the
lowest/highest molecular weight, and 1--3 other randomly chosen members
of the cluster. 3DX supported multiple sessions, so an expert could
stop and resume the evaluation when convenient. The expert could
evaluate the clusters in the subset in any order, but a new random
subset of clusters, excluding the ones already rated, was assigned for
evaluation only when all the clusters in the previous subset had been
evaluated or when the expert resumed the evaluation after interrupting
the previous session for a break. Clusters assigned but not evaluated
could, in principle, be assigned again in another session.
Interestingly, some experts rated all compounds, for which they took a
considerable amount of time, which is necessary to avoid jeopardizing
face-validity.

\begin{figure}

\includegraphics{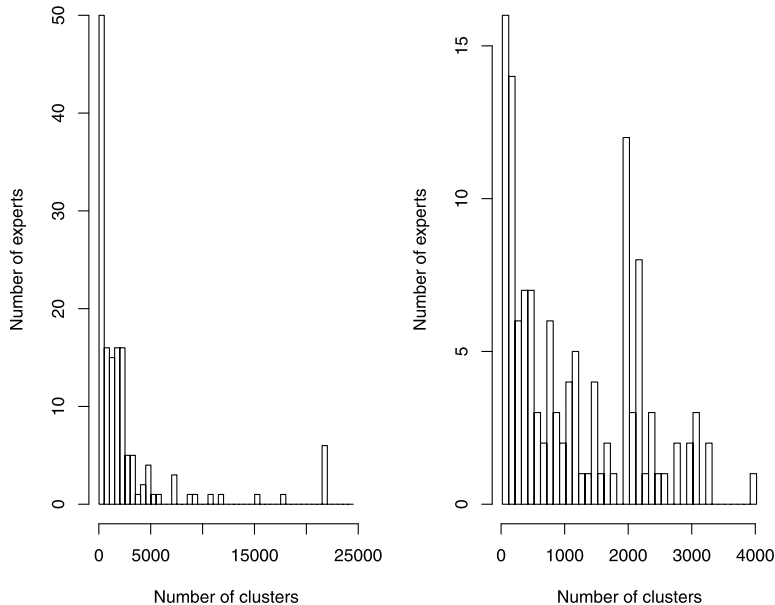}

\caption{Histograms of the number of clusters evaluated by the
experts: The left panel displays the
information from all experts, and the right panel displays the
information for experts who
evaluated fewer than 4000 clusters.} \label{fig:compounds-dist}
\end{figure}

The histogram in the left panel of Figure~\ref{fig:compounds-dist}
displays the distribution of the number of clusters evaluated by the
experts. As one would expect, many experts opted to evaluate a
relatively small number of clusters. Indeed, $25\%$ of the experts
evaluated fewer than 345 clusters, $50\%$ fewer than 1200, and $75\%$
fewer than 2370 clusters. The right panel displays the distribution for
those experts who evaluated fewer than 4000 clusters. It confirms that
experts tended to evaluate only a small percentage of all the clusters
and has notable peaks at 0--200 and 2000. In total, the final data set
contained 409,552 observations.

\subsection{High-dimensional data}\label{sec:intro}
Steady advances in fields like genetics and molecular biology are
dramatically increasing our capacity to create chemical compounds for
therapeutic use. Nevertheless, developing these compounds into
effective drugs is an expensive and lengthy process, and consequently
pharmaceutical companies need to carefully evaluate their potential
before investing more resources. Expert opinion has been acknowledged
as a crucial element in this evaluation process [\citet{Andy}, \citet{Hack2011}]. In practice, similar
compounds are grouped into clusters whose potential is qualitatively
assessed by experts. We show that, using these qualitative assessments
and hierarchical models, a probability of \emph{success} can be
assigned to each cluster, where success entails recommending the
inclusion of a cluster in the sponsor's database for future scrutiny.
However, the presence of several experts and many clusters leads to a
high-dimensional vector of repeated responses and fixed effects,
creating a serious computational challenge.

Facets of the so-called \emph{curse of dimensionality} are numerous in
statistics and constitute active areas of research [\citet{dono00,fan2006}]. \citet{T96} studied regression shrinkage and selection via
the lasso; his paper is an excellent example of the need for and
popularity of methods for high-dimensional data. \citet{fews} proposed
several approaches to fit multivariate hierarchical models in settings
where the responses are high-dimensional vectors of repeated observations.

\citet{xia} categorized methods that deal with high
dimensionality into data reduction and functional approaches [\citet{li,rich}]. Following the data reduction route, we propose a method
that circumvents the problem of dimensionality and allows a reliable
assessment of the probability of success for each cluster. The approach
is based on permuting and splitting the original data set into mutually
exclusive subsets that are analyzed separately and the posterior
combination of the results from these analyses. It aims to render the
use of random-effects models possible when the data involve a huge
number of clusters and/or a large number of experts.

Data-splitting methods have already been used for tackling
high-\break dimensional problems. For instance, \citet{chen12} used a
split-and-conquer approach to analyze extraordinarily large data in
penalized regression, \citet{fan12} employed a data-splitting technique
to estimate the variance in ultrahigh-dimensional linear regression,
and \citet{molsplit} formulated a splitting approach for model fitting
when either the repeated response vector is high-dimensional or the
sample size is too large.

Nonetheless, the scenario studied in this paper is radically different
because both the response vector and the vector of fixed effects are
high dimensional. This structure requires a splitting strategy in which
the parameters and Hessian matrices estimated in each subsample are not
the same and, therefore, the methods mentioned above do not directly apply.

The paper is organized as follows. Section~\ref{sec:method} introduces
the methodology mentioned above. Section~\ref{sec:results} discusses
results from applying the methodology to the case study. To assess the
performance of the new approach, we carried out a simulation study.
Section~\ref{sec:simdesign} outlines its design and main findings.
Section~\ref{sec:conclusion} gives some final comments and conclusions.

\section{Estimating the probability of success}\label{sec:method}

To facilitate the decision-\break making process, it is desirable to summarize
the qualitative assessments in a single probability of success for each
cluster. One approach uses generalized linear mixed models. A simpler
method uses the observed probabilities of success, estimated as the
proportion of 1's that each cluster received. There are, however, good
reasons to prefer the model-based approach. Hierarchical models can
include covariates associated with the clusters and the experts. They
also permit extensions to
compensate for selection bias or missing data and explicitly account
for an expert's evaluation of several clusters. In addition, the
model-based approach naturally delivers an estimate of the inter-expert
variability. Although it is not the focus of the analysis, a measure of
heterogeneity among experts is valuable for interpretation of the
results and for design of future evaluation studies.

To estimate the probability of success for each of the $N$ clusters, we
denote the vector of ratings associated with expert $i$ by $\Y_i =
(Y_{ij})_{j\in\Lambda_i}$, where $\Lambda_i$ is the set of clusters
evaluated by expert $i$ ($i=1,\dots,n$). A natural choice is the
logistic-normal model
%
\begin{equation}
\label{eq:modelmain} \operatorname{logit} \bigl[P (Y_{ij}=1|\beta_j,b_i
) \bigr]=\beta_j + b_i,
\end{equation}
where $\beta_j$ is a fixed effect for cluster $C_{j}$ with $j \in
\varLambda_i$ and ${  b_{i} \sim N(0,\sigma^2)}$ for expert
$i$ is a random effect. Models similar to (\ref{eq:modelmain}) have
been successfully applied in psychometrics to describe the ratings of
individuals on the items of a test or psychiatric scale. In that
context, model (\ref{eq:modelmain}) is known as the Rasch model and
plays an important role in conceptualization of fundamental measurement
in psychology, psychiatry and educational testing [\citet{boeck2,Bond2007}].
The problem studied in this work has clear similarities with the
measurement problem in psychometrics. For instance, the clusters in our
setting parallel the items in a test or psychiatric scale, and the
ratings of an individual on these items would be equivalent to the
ratings given by the experts in our setting. Nonetheless, differences
in the
target of inference and the dimension of the parameter space imply that
the two areas need distinct approaches.

Parameter estimates for model (\ref{eq:modelmain}) are obtained by
maximizing the likelihood,
%
\begin{equation}
\label{eq:lik} L\bigl(\bbeta, \sigma^2\bigr)=\prod
_{i=1}^{n} \int^{\infty}_{-\infty}{
\prod_{{j
\in\varLambda_i}}}\pi_{ij}^{y_{ij}}(1-
\pi_{ij})^{1-y_{ij}} \phi\bigl(b_i | 0,
\sigma^2\bigr) \,db_i,
\end{equation}
using, for example, a Newton--Raphson optimization algorithm, where $\pi
_{ij}=P (Y_{ij}=1|\beta_j,b_i )$, $\bbeta=(\beta_1,\ldots,\beta
_N)'$ contains the cluster effects and $\phi(b_i | 0,\sigma^2)$ denotes
the normal density with mean 0 and variance $\sigma^2$. The integral
can be approximated by applying numerical procedures such as
Gauss--Hermite quadrature.

Using model (\ref{eq:modelmain}), one can calculate the marginal
probability of success for cluster $C_j$ by integrating over the
distribution of the random effects
%
\begin{equation}
\label{eq:marginal-P} P_j=P \bigl(Y_{j}=1|\beta_j,
\sigma^2 \bigr)=\int\frac{\exp (\beta
_j + b  )}{1+\exp (\beta_j + b  )} \phi\bigl(b | 0,\sigma
^2\bigr) \,db.
\end{equation}
One first estimates the cluster effects $\beta_j$, after adjusting for
the expert effects, by maximizing the likelihood (\ref{eq:lik}). One
then uses these estimates to estimate the probability of success by
averaging over the entire population of experts. However, the vector of
fixed effects $\bbeta$ in (\ref{eq:lik}) has dimension 22,015, and the
dimension of the response vector $\Y_i$ ranges from 20 to 22,015.
Hence, maximum likelihood is not feasible with the most commonly
available computing resources.
In particular, Gauss--Hermite or other quadrature methods, used to
evaluate the integrals in (\ref{eq:marginal-P}), can be particularly
challenging [\citet{PB95,mol05}].
The challenge is then to find a reasonable strategy for estimating the
probabilities of interest. Alternatively, one may consider stochastic
integration instead, as we do below.

\subsection{A permutational-splitting sample procedure}

Let ${\cal C}= \{{\cal C}_1,\dots, {\cal C}_N \}$ denote the
collection of ratings on the $N$ clusters, where ${\cal C}_j$ is a
vector containing all the ratings cluster $C_j$ received. Our procedure
partitions the set of cluster evaluations ${\cal C}$ into $S$ disjoint
subsets of relatively small size. As with any splitting procedure, one
must decide on the size of these subsets. In our setting, if $N_k$
denotes the number of vectors ${\cal C}_j$ in subset $k$ (where $ N_1 +
N_2 + \cdots+ N_S=N$), then one needs to determine the $N_k$ so that
model (\ref{eq:modelmain}) can be fitted, with commonly available
computing resources, using maximum likelihood and the information in
each subset. Even though the search for appropriate $N_k$ may produce
more than one plausible choice, a sensitivity analysis could easily
explore the impact of these choices on the conclusions. For instance,
in our case study,
$N_k=15$ and $N_k=30$ gave very similar results, indicating a degree of
robustness to the choice of $N_k$. In general, the subsets' size may
vary from one application to another. However, 30--40 clusters per
subset seem to be a reasonable starting point. Clearly, the choice of
the $N_k$ determines $S$, and some subsets may have slightly more or
fewer clusters than $N_k$ when $N/N_k$ is not a whole number.
Taking these ideas into account, we developed the following procedure:
\begin{longlist}[1.]
\item[1.] \textit{Splitting}. 
Split the set ${\cal C}$ into $S$ mutually exclusive and exhaustive
subsets ${\cal C}^k$ (${  k=1,\ldots,S}$) with $N_k<N$
denoting the number of clusters in ${\cal C}^k$. The information in
these subsets may not be independent because ratings from the same
expert may appear in more than one subset. However, because the subsets
are exclusive and exhaustive,
a given cluster belongs to a single subset.
\item[2.]\textit{Estimation}. 
Using maximum likelihood and the
information included in each~${\cal C}^k$, fit model (\ref
{eq:modelmain}) $S$ times. For all $k$, $N_k<N$ (typically $N_k\ll N$),
so the dimensions of the response and fixed-effect vectors in these
models are much smaller. Merging all estimates obtained from these
fittings leads to an estimate of the vector of fixed-effect parameters
and $S$ estimates of the random-effect variance $\sigma^2$. Clearly,
within each subset, the estimator of the inter-expert variance $\widehat
{\sigma}_k^2$ uses information from only a subgroup of all experts and
thus is less efficient than the estimator based on all data. The
pooling of the subset-specific estimates should not be done
mechanically; a careful analysis should look for unusual behavior. The
procedure described in the next step may help in checking the stability
of the parameter estimates.

\item[3.] \textit{Permutation}. 
Randomly permute the elements of
${\cal C}$, and repeat steps 1 and~2 $W$ times.
This step is equivalent to sampling without replacement from the set of
all possible partitions introduced in step 1. Consequently,
instead of estimating the parameters of interest based on a single
arbitrary partition, their estimation is based on multiple randomly
selected partitions of the set of clusters. The permutation step serves
several purposes. It yields estimates of the parameters based on
different subsamples of the same data and, hence, makes it possible to
check the stability of the estimates. This diversity may be especially
relevant for the variance component, because it is estimated with
multiple sample sizes. In addition, combining estimates from different
subsamples produces more reliable final estimates. To capitalize on
these features, one should ideally consider a large number of
permutations ($W$). Our results, however, indicate little gain from
taking $W$ larger than~$20$.

\item[4.] \textit{Estimation of the success probabilities}. 
Step 3 produces the estimates $\widehat{\bbeta}_w$ and $\widehat
{\sigma}_{kw}^2$, where $w=1,\ldots,W$ and $k=1,\ldots,S$.
Subsequently, based on $\widehat{\bbeta}_w$ and $\widehat{\sigma
}_w^2=\frac{1}{S}\sum^{S}_{k=1}{\widehat{\sigma}_{kw}^2}$, estimates of
the success probability of each cluster can be obtained using (\ref
{eq:marginal-P}), with the integral computed via stochastic integration
by drawing $Q$ elements $b_q$ from $N(0,\widehat{\sigma}_w^2)$.
Importantly, unlike $\widehat{\sigma}_{kw}^2$, which only uses
information from the experts in subset $k$, $\widehat{\sigma}_w^2$ is
based on information from all experts and hence offers a better
assessment of the inter-expert variance. It is of course possible, when
needed, to optimize this stochastic procedure. Eventually, the
probability of success for cluster $C_j$ can be estimated as
\[
\widehat{P}_{j}=\frac{1}{W}\sum^{W}_{w=1}
\widehat{P}_{wj}\qquad \mbox {where } \widehat{P}_{wj}=
\widehat{P}_{w} (Y_{j}=1 )=\frac{1}{Q}\sum
^{Q}_{q=1}\frac{\exp (\widehat{\beta}_{wj} + b_q  )}{1+\exp
 (\widehat{\beta}_{wj} + b_q  )}.
\]
Similarly,
\[
\widehat{\beta}_{j}=\frac{1}{W}\sum
^{W}_{w=1}\widehat{\beta }_{wj}\quad \mbox{and}\quad
\widehat{\sigma}^2=\frac{1}{W}\sum
^{W}_{w=1}{\widehat{\sigma}_{w}^2}.
\]

One may heuristically argue that step~3 also ensures that
final estimates of the cluster effects are similar to those obtained if
maximum likelihood were used with the whole data. Indeed, let $\widehat
{\beta}_{wj}$ denote again the maximum likelihood estimators for the
effect of cluster $C_j$ computed in each of the $W$ permutations and
$\widehat{\beta}_{Nj}$ the maximum likelihood estimator based on the
entire set of $N$ clusters. Further, consider the expression $\widehat
{\beta}_{wj}=\widehat{\beta}_{Nj} +e_{wj}$,\vspace*{1pt} where $e_{wj}$ is the
random component by which $\widehat{\beta}_{wj}$ differs from $\widehat
{\beta}_{Nj}$. Because maximum likelihood estimators are asymptotically
unbiased, provided maximum likelihood is estimating the same
parameters, one has $E(e_{wj})\approx0$; and extensions of the law of
large numbers for correlated, not identically distributed random
variables, may suggest that, under certain assumptions, for a
sufficiently large $W$ [\citet{Newman1984,Birkel1992}]
\[
\widehat{\beta}_{j}= \frac{1}{W}\sum
^{W}_{w=1}{\widehat {\beta}_{wj}}=\widehat{
\beta}_{Nj} + \frac{1}{W}\sum^{W}_{w=1}{e_{wj}}
\approx\widehat{\beta}_{Nj}.
\]
Similar arguments apply to the variance component and the success
probabilities. The findings of the simulation study presented in
Section~\ref{sec:simdesign} support these heuristic results. In a
particular data set, this argument could further be verified by
comparing the split procedure with full maximum likelihood. When the
latter is not feasible, one could consider a subset for which full
likelihood is feasible. Of course, when chosen too small, the
discrepancy between the two procedures could well be considerably
larger than what it is for the entire set of data.

\item[5.] \textit{Confidence intervals for the success probabilities}. 
To construct a confidence interval for the success
probability of cluster $C_j$, we consider the results from one of the
$W$ permutations described in step 3. To simplify notation,
we omit the subscript $w$, but these calculations are meant to be done
for each of the $W$ permutations.

If ${\cal C}^k$ denotes the unique subset of ${\cal C}$ containing
${\cal C}_j$, then fitting model (\ref{eq:modelmain}) to ${\cal C}^k$
produces the maximum likelihood estimator $\widehat{\btheta
}_{j}=(\widehat{\beta}_{j}, \widehat{\sigma}^2_{k})'$. Classical
likelihood theory guarantees that, asymptotically, $\widehat{\btheta
}_{j}\sim N  (\btheta_{j}, \bSigma )$, where a consistent
estimator of the $2\times2$ matrix $\bSigma$ can be constructed using
the Hessian matrix obtained from fitting the model. Even though the
estimator $\widehat{\sigma}^2_{k}$ is not efficient, its use is
necessary in this case to directly apply asymptotic results from
maximum likelihood theory. For a sufficiently large value of $W$, one
could derive a confidence interval for each $P_j$, based on replication.

The success probability $P_{j}$ is a function of $\btheta_{j}$, such
that if one defines $\gamma_{j}=\log \{{P_{j}}/({1-P_{j}}) \}
$, then the delta method leads to ${ \widehat{\gamma
}_{j}\sim N  (\gamma_{j}, \sigma_{\gamma}^2  )}$
asymptotically, where $\widehat{\gamma}_{j}=\log \{{\widehat
{P}_{j}}/({1-\widehat{P}_{j}}) \}$ and
\begin{eqnarray*}
\sigma_{\gamma}^2&=& \biggl(\frac{\partial{\gamma}_{j}}{\partial{\btheta
_{j}}} \biggr) \bSigma
\biggl(\frac{\partial{\gamma}_{j}}{\partial
{\btheta_{j}}} \biggr)^{\prime},
\\
\frac{\partial{\gamma}_{j}}{\partial{\btheta_{j}}}&=&\frac
{1}{P_{j}(1-P_{j})}\frac{\partial{P}_{j}}{\partial{\btheta_{j}}},
\end{eqnarray*}
with
\begin{eqnarray*}
\frac{\partial{P}_{j}}{\partial{\beta_{j}}}&=&\int\frac{\exp (\beta
_{j} + b  )}{ \{1+\exp (\beta_{j} + b  ) \}
^2} \phi\bigl(b | 0,\sigma^2_{k}
\bigr) \,db,
\\
\frac{\partial{P}_{j}}{\partial{\sigma^2_{k}}}&=&\int\frac{\exp
(\beta_{j} + b  )}{1+\exp (\beta_{j} + b  )} \frac
{b^2-\sigma^2_{k}}{2\sigma^4_{k}} \phi\bigl(b| 0,
\sigma^2_{k}\bigr) \,db.
\end{eqnarray*}
The necessary estimates can be obtained by plugging $\widehat{\btheta
}_{j}$ into the corresponding expressions and using stochastic
integration as previously described. Finally, an asymptotic $95\%$
confidence interval for $P_{j}$ is given by
\[
\mathrm{CI}_{P_{j}}=\frac{\exp (\widehat{\gamma}_{j}\pm1.96\cdot\widehat
{\sigma}_{\gamma} )}{1+\exp (\widehat{\gamma}_{j}\pm1.96\cdot
\widehat{\sigma}_{\gamma} )}.
\]
The overall confidence interval follows from averaging the lower and
upper bounds of all confidence intervals from the $W$ partitions. A
more conservative approach would consider the minimum of the lower
bounds and the maximum of the upper bounds, that is, the union
interval. In reverse, the intersection interval (maximum of the lower
bounds; minimum of the upper bounds) might be too liberal. In
principle, one should adjust the coverage probabilities using, for
example, a Bonferroni correction when constructing these intervals. If
the overall coverage probability for the entire family of confidence
intervals is $95\%$, then it is easy to show that the overall
confidence interval will have a coverage probability of at least $95\%
$. This implies construction of confidence intervals with level
($1-0.05/W$) for $P_j$ in each permutation, which are likely to be too
wide for useful inference. In Section~\ref{sec:simdesign} we study the
performance of this interval via simulation without using any
correction, and the results confirm that in many practical situations
this simpler approach may work well. Of course, the resulting interval
is then for a single $P_j$. In case simultaneous inference for several
$P_j$ is needed, conventional adjustments need to be made.
\end{longlist}

In these developments, we assume that, \textit{given cluster and expert
effects}, an expert's evaluations of different clusters are
independent. The correctness of this assumption is relevant when
different clusters, evaluated by the same expert, end up in the same
partitioning set. Our assumption is similar to the psychometric
assumption that items' difficulties are intrinsic characteristics. Even
though we believe that this assumption is reasonable, it is
nevertheless important to be aware of it.

%
\begin{table}
\caption{The $20$ clusters (ID) with highest
estimated probability of success: Estimated cluster effect ($\widehat{\beta
}_j$), Estimated/Observed success probabilities (proportion of 1's for
each cluster) and confidence interval limits}
\label{dich1}
\begin{tabular*}{\textwidth}{@{\extracolsep{\fill}}lccccc@{}}
\hline
&&\multicolumn{2}{c}{\textbf{Probability}}&
\multicolumn{2}{c@{}}{\textbf{95\% CI}}\\[-6pt]
&&\multicolumn{2}{c}{\hrulefill}&\multicolumn{2}{c@{}}{\hrulefill}\\
\multicolumn{1}{@{}l}{\textbf{ID}}&$\bolds{\widehat{\beta}_j}$&\textbf{Estimated}&\textbf{Observed}&\textbf{Lower}&\textbf{Upper}\\
\hline
295061 & 3.07& 0.80& 0.82& 0.58 & 0.92\\
296535 & 2.51& 0.76& 0.81& 0.51 & 0.90\\
\phantom{0}84163 & 2.40& 0.75& 0.78& 0.48 & 0.90\\
313914 & 2.30& 0.74& 0.80& 0.39 & 0.93\\
265441 & 2.16& 0.72& 0.69& 0.50 & 0.87\\
296443 & 2.09& 0.72& 0.62& 0.52 & 0.86\\
277774 & 2.01& 0.71& 0.71& 0.49 & 0.86\\
265222 & 1.96& 0.71& 0.70& 0.53 & 0.84\\
178994 & 1.84& 0.69& 0.73& 0.50 & 0.84\\
462994 & 1.73& 0.69& 0.69& 0.44 & 0.86\\
292579 & 1.76& 0.69& 0.75& 0.45 & 0.84\\
296560 & 1.71& 0.68& 0.72& 0.47 & 0.83\\
277619 & 1.67& 0.68& 0.63& 0.47 & 0.83\\
315928 & 1.67& 0.68& 0.75& 0.47 & 0.84\\
296427 & 1.69& 0.68& 0.78& 0.35 & 0.91\\
263047 & 1.60& 0.68& 0.76& 0.45 & 0.84\\
333529 & 1.62& 0.67& 0.80& 0.45 & 0.84\\
292805 & 1.52& 0.67& 0.72& 0.43 & 0.85\\
178828 & 1.43& 0.66& 0.72& 0.43 & 0.83\\
265229 & 1.39& 0.65& 0.65& 0.47 & 0.80\\[3pt]
$\widehat{\sigma}^2$&\multicolumn{1}{c}{10.279}&&&&\\
\hline
\end{tabular*}
\end{table}

\section{Data analysis}\label{sec:results}
\subsection{Unweighted analysis}
The procedure introduced in Section~\ref{sec:method} was applied to the
data described in Section~\ref{sec:case}, using $N_k=30$, $Q=10\mbox{,}000$,
$S=734$ and $W=20$. Table~\ref{dich1} gives the results for the 20
top-ranked clusters, that is, the clusters with the highest estimated
probability of success. All clusters in the table have an estimated
probability larger than 60\%, and the top 3 have probability of success
around 75\%. The observed probabilities (proportion of 1's for each
cluster) lie within the 95\% confidence limits of their corresponding
model-based probability estimates. In spite of this, reasonable
differences, close to 0.1, are observed for some clusters (e.g.,
296443, 296427 and 333529) and this may signal a potential problem in
regard to the use of observed probabilities. Importantly, these naive
estimates completely ignore the correlation between ratings from the
same expert. Therefore, they do not correct for the possibility that
some experts may tend to give higher/lower ratings than others and may
lead to biased estimates for clusters that are mostly evaluated by
definite/skeptical experts. In addition, the results indicate high
heterogeneity among experts, with estimated variance
\[
{ \widehat{\sigma}^2=\frac{1}{W}\sum
^{W}_{w=1}{\widehat {\sigma}_{w}^2}
\approx10}.
\]
On the one hand, this large variance may indicate a need to select
experts from a more uniform population by applying, for example, more
stringent selection criteria. On the other hand, more stringent
selection criteria may conflict with having experts that represent an
appropriately broad range of opinions. In this sense, a broad range may
be considered beneficial, provided the model used properly accommodates
between-expert variability. Finding a balance between these two
considerations is very important for the overall quality of the study.
In general, if experts show substantial heterogeneity, then additional
investigation should try to determine the source before further actions
are taken.

In principle, it is possible to use fixed effects for the 147 experts.
Of course, this would raise the issue of inconsistency when the number
of experts increases. Apart from this, the estimated fixed effects
could be examined informally to assess heterogeneity in the sample of raters.

\begin{figure}

\includegraphics{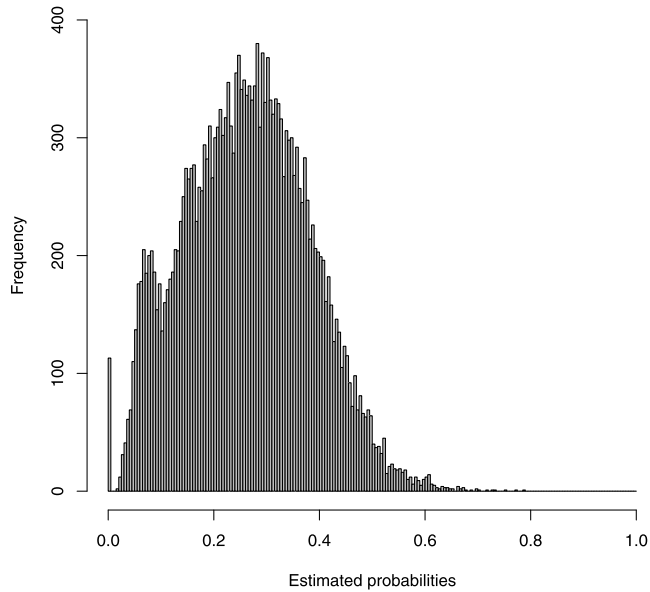}

\caption{Distribution of estimated probabilities of success.}\label
{predprob-dist}
\end{figure}

The general behavior of the estimated probabilities of success is
displayed in Figure~\ref{predprob-dist}. Visibly, most clusters have a
quite low probability of success, with the median around 26\%, and 75\%
of the clusters have an estimated probability of success smaller than
40\%. About 120 clusters are unanimously not recommended, as evidenced
by the peak at zero probability. This is in line with the observed
data: none of them received a positive recommendation, though their
numbers of evaluations ranged between $11$ and $23$. Another
conspicuous group contains clusters that had only 1--3 positive
evaluations and, as expected, produced low estimated proportions of
success, ranging between 8 and 10\%.

The interpretation of these probabilities will frequently be
subject-specific. Taking into account the economic cost associated with
the development of these clusters, the time frame required to develop
them, and the potential social and economic gains that they may bring,
researchers can define the minimum probability of success that would
justify further study.

The analysis of the confidence intervals also offers some important
insight. First, although moderately wide, the confidence intervals
still allow useful inferences. The large inter-expert heterogeneity may
hint at possible measures to increase precision in future studies.
Second, using the lower bound of the confidence intervals to rank the
clusters, instead of the point estimate of the probability of success,
may yield different results. By this criterion, cluster 265222, ranked
eighth by the point estimate, would become the second most promising
candidate. Clearly, some more fundamental, substantive considerations
may be needed to complement the information in Table~\ref{dich1} during
the decision-making process.

As a sensitivity analysis we also considered $N_k=15$, $W=20$,
$S=1468$, with $Q=10\mbox{,}000$. The results appear in the columns labeled
``unweighted'' in Table~\ref{weightvsnoweight}. Clearly, the
differences with the original analysis are negligible.

\begin{table}
\caption{Estimates for the fixed effects
and probabilities of success obtained from the weighted and unweighted
analyses for the top $20$ clusters in terms of unweighted probability;
$\widehat{\beta}_{\mathrm{weighted}}$ and $\widehat{\beta}_{\mathrm{unweighted}}$
are the estimated cluster effects with the ranks in parentheses, and
$\widehat{\operatorname{prob}}_{\mathrm{weighted}}$ and $\widehat{\operatorname
{prob}}_{\mathrm{unweighted}}$ are the corresponding probabilities of success}
\label{weightvsnoweight}
\begin{tabular*}{\textwidth}{@{\extracolsep{\fill}}ld{2.2}ccccc@{}}
\hline
\multicolumn{1}{@{}l}{\textbf{ID}}&
\multicolumn{1}{c}{$\bolds{\widehat{\beta}_{\mathrm{weighted}}}$}&
\multicolumn{1}{c}{$\bolds{\widehat{\beta}_{\mathrm{{unweighted}}}}$}&
\multicolumn{1}{c}{$\bolds{\widehat{\mathrm{prob}}_{\mathrm {weighted}}}$}&&
\multicolumn{1}{c}{$\bolds{\widehat{\mathrm{prob}}_{\mathrm{unweighted}}}$}&\\
\hline
295061& 3.86& 3.33& 0.90 & (2) & 0.80 & (1) \\
296535& 1.99& 2.71& 0.74 & (54) & 0.76 & (2) \\
\phantom{0}84163& 0.86& 2.42& 0.61 & (376) & 0.73 & (3) \\
296443& 0.54& 2.41& 0.57 & (620) & 0.73 & (4) \\
313914& 3.79& 2.37& 0.89 & (3) & 0.73 & (5) \\
265222& 0.56& 2.40& 0.57 & (653) & 0.73 & (6) \\
333529& 1.85& 1.99& 0.73 & (67) & 0.69 & (7) \\
296560& 1.26& 1.91& 0.66 & (198) & 0.69 & (8) \\
178994& 2.25& 1.91& 0.77 & (28) & 0.69 & (9) \\
265441& 1.22& 1.94& 0.66 & (211) & 0.69 & (10) \\
277774& 2.26& 1.87& 0.77 & (29) & 0.69 & (11) \\
292579& 2.69& 1.91& 0.81 & (10) & 0.69 & (12) \\
315928& 1.18& 1.87& 0.65 & (233) & 0.68 & (13) \\
277619& -0.63& 1.74& 0.42 & (3165) & 0.67 & (14) \\
263047& 3.85& 1.78& 0.90 & (1) & 0.67 & (15) \\
296427& 2.70& 1.65& 0.81 & (12) & 0.67 & (16) \\
292805& 1.00& 1.60& 0.63 & (313) & 0.66 & (17) \\
178828& 2.26& 1.52& 0.77 & (27) & 0.66 & (18) \\
462994& 1.31& 1.46& 0.67 & (183) & 0.65 & (19) \\
159643& 1.93& 1.50& 0.74 & (55) & 0.65 & (20) \\[3pt]
$\widehat{\sigma}^2$&3.19&15.80&&&&\\
\hline
\end{tabular*}\vspace*{-3pt}
\end{table}
\subsection{Weighted analysis}
An important issue discussed in Section~\ref{sec:case} was the
differences encountered in the numbers of clusters evaluated by the
experts. One may wonder whether experts who evaluated a large number of
clusters gave as careful consideration to each cluster as those who
evaluated only a few. Importantly, the model-based approach introduced
in Section~\ref{sec:method} can take into account these differences by
carrying out a weighted analysis, which maximizes the likelihood function
%
\begin{equation}
\label{lik-wgt} L\bigl(\bbeta, \sigma^2\bigr)=\prod
_{i=1}^{n} \omega_i\int
^{\infty}_{-\infty
}{\prod_{{j \in\varLambda_i}}}
\pi_{ij}^{y_{ij}}(1-\pi_{ij})^{1-y_{ij}} \phi
\bigl(b_i | 0,\sigma^2\bigr) \,db_i,
\end{equation}
where $\omega_i=N/|\varLambda_i|$ and $|\varLambda_i|$ denotes the size
of $\varLambda_i$. Practically, a weighted analysis, using the SAS
procedure NLMIXED, implies replication of each response vector by
$\omega_i$, resulting in a pseudo-data set with larger sample size than
in the unweighted analysis. Using partitions with $N_k=30$ was rather
challenging; consequently, the weighted analysis was carried out with
$N_k=15$. The main results are displayed in Table~\ref{weightvsnoweight}.

Interestingly, some important differences emerge from the two
approaches. For instance, the top-ranked cluster in the unweighted
analysis received rank 2 in the weighted approach. Some differences are
even more dramatic; for example, the fourth cluster in the unweighted
analysis received rank 620 in the weighted approach. Clearly, a very
careful and thoughtful discussion of these differences will be needed
during the decision-making process.\vadjust{\goodbreak} In addition, these results also
point out the importance of a careful design of the study and may
suggest changes in the design to avoid large differences in the numbers
of clusters evaluated by the experts. The top 20 in Table~\ref{weightvsnoweight} is very
similar to the one in Table~\ref{dich1}, but it is not exactly the
same. For example, the cluster ranked 20th is not in Table~\ref{dich1},
probably because of the change in $\widehat{\sigma}^2$.

Fitting model (\ref{eq:modelmain}) to the entire data set using maximum
likelihood was unfeasible in this case study. Therefore, all previous
conclusions were derived by implementing the procedure described in
Section~\ref{sec:method}. One may wonder how the previous procedure
would compare with maximum likelihood when the latter is tractable. In
the next section we investigate this important issue via simulation.

\section{Simulation study}\label{sec:simdesign}
The simulations were designed to mimic the main characteristics
encountered in the case study. Two hundred data sets were generated,
with the following parameters held constant across data sets: (1)
Number of clusters $N=50$, chosen to ensure tractability of maximum
likelihood estimation for the whole data, (2) number of experts
$n=147$, and (3) a set of 50 values assigned to the parameters
characterizing the cluster effects ($\beta_j$), which were sampled from
a $N(-2,2)$ one time and then held constant in all data sets. Factors
varying among the data sets were as follows: (1) the number of ratings
per expert $n_i$, independently sampled from $\operatorname{Poisson}(25)$ and restricted
to the range of 8 to 50, and (2) a set of $147$ expert random effects
$b_i$, independently sampled from $N(0,12.25$). Conceptually, each
generated data set represents a replication of the evaluation study in
which a new set of experts rates the same clusters. Therefore, varying
$b_i$ from one data set to another resembles the use of different
groups of experts in each study, sampled from the entire population of
experts. Clearly, $n_i$ needs to vary simultaneously with $b_i$. The
probability that expert $i$ would recommend the inclusion of cluster
$j$ in the sponsor's database, $P_{ij}=P(Y_{ij}=1|\beta_j, b_i)$, was
computed using model (\ref{eq:modelmain}) and the response $Y_{ij} \sim
\operatorname{Bernoulli}(P_{ij})$. Finally, model (\ref{eq:modelmain}) was
fitted using full maximum likelihood and the procedure introduced in
Section~\ref{sec:method}, and their corresponding probabilities of
success, given by (\ref{eq:marginal-P}), were compared. Parameters used
in the split procedure were $N_k=5$, $W=20$, $Q=10\mbox{,}000$ and $S=10$.

\begin{figure}

\includegraphics{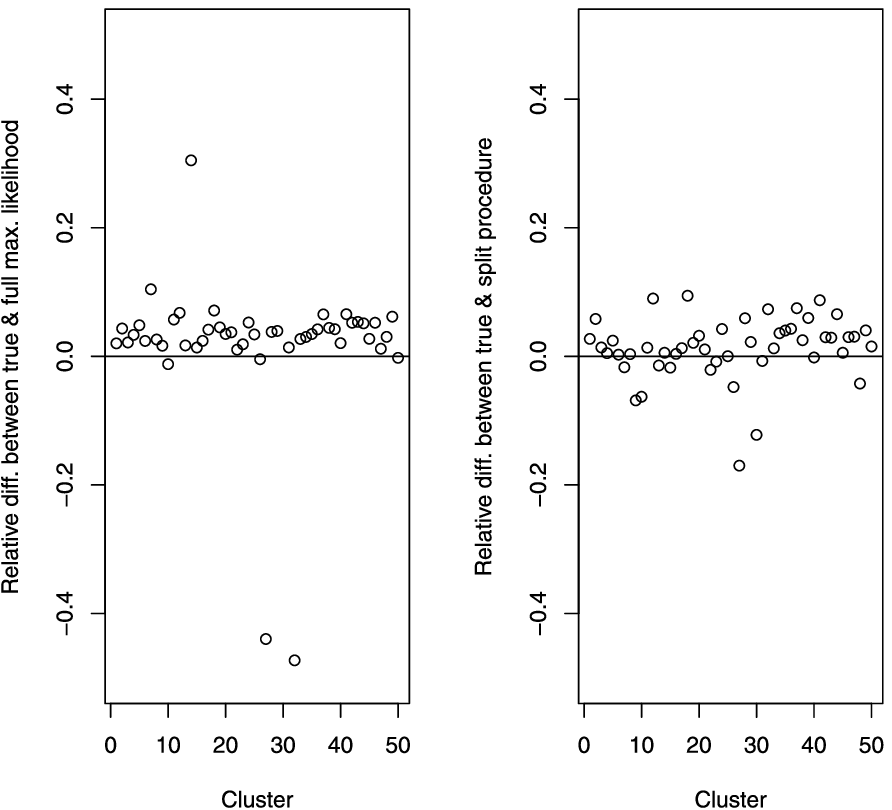}

\caption{Average relative difference between the true values and
the estimates obtained from maximum likelihood,
$\frac{ \widehat{\beta}_{j,\mathrm{mle}}-\beta_j}{\beta_j}$ (left) and the
split procedure
$\frac{ \widehat{\beta}_{j,\mathrm{split}}-\beta_j}{\beta_j}$(right). (Results
of the simulation study.)}
\label{differences}
\end{figure}

The main results of the simulation study for the top $20$ clusters
(those with the highest true probability of success) are summarized in
Tables~\ref{simestimates} and \ref{simprobabilities}. Table~\ref{simestimates} clearly shows that the proposed procedure performs as
well as maximum likelihood, for the point estimates of the cluster
effect. Further, Figure~\ref{differences} shows that this is true for
most of the clusters, as the average relative differences from the true
values, for the maximum likelihood estimators [$(\widehat{\beta
}_{j,\mathrm{mle}}-\beta_j )/{\beta_j}$] and the estimators obtained from the
split procedure [$(\widehat{\beta}_{j,\mathrm{split}}-\beta_j )/{\beta_j}$], are
very close to zero most of the time.\vadjust{\goodbreak}
Interestingly, maximum likelihood cluster-effect estimates for clusters
14, 27, 30 and 32 have a noticeably larger average relative bias than
their split-procedure counterparts (\#30 is off the scale). This
results from the fact that, for these four clusters, the denominator in
the relative-difference expression is very small, highlighting a
well-known shortcoming of ratios and relative differences.
In Table~\ref{simprobabilities}, the corresponding values are unremarkable.
%
\begin{table}
\caption{True value and average parameter
estimate for the top $20$ clusters (ID) in the simulation study,
estimated from full maximum likelihood (likelihood) and the split
procedure (procedure)}
\label{simestimates}
\begin{tabular*}{\textwidth}{@{\extracolsep{\fill}}ld{2.2}d{2.2}d{2.2}@{}}
\hline
&\multicolumn{3}{c@{}}{$\bolds{\beta_j}$}\\[-6pt]
&\multicolumn{3}{c@{}}{\hrulefill}\\
\multicolumn{1}{@{}l}{\textbf{ID}}& \multicolumn{1}{c}{\textbf{True}} &\multicolumn
{1}{c}{\textbf{Likelihood}}&\multicolumn{1}{c@{}}{\textbf{Procedure}}\\
\hline
\phantom{0}3 & 2.33 & 2.38 & 2.36 \\
\phantom{0}1 & 1.60 & 1.63 & 1.65 \\
33 & 1.52 & 1.56 & 1.54 \\
47 & 1.43 & 1.45 & 1.48 \\
50 & 1.04 & 1.03 & 1.05 \\
27 & 0.13 & 0.07 & 0.11 \\
30 & 0.06 & 0.01 & 0.05 \\
32 & 0.06 & 0.03 & 0.06 \\
14 & -0.11 & -0.14 & -0.11 \\
\phantom{0}7 & -0.30 & -0.33 & -0.29 \\
\phantom{0}9 & -0.49 & -0.50 & -0.46 \\
48 & -0.63 & -0.65 & -0.61 \\
10 & -0.71 & -0.70 & -0.66 \\
21 & -0.97 & -1.00 & -0.98 \\
11 & -1.12 & -1.19 & -1.14 \\
26 & -1.13 & -1.12 & -1.07 \\
15 & -1.32 & -1.33 & -1.29 \\
13 & -1.40 & -1.42 & -1.38 \\
\phantom{0}4 & -1.42 & -1.47 & -1.42 \\
42 & -1.61 & -1.69 & -1.66 \\[3pt]
$\widehat{\sigma}^2$&12.25&12.96& 12.74\\
\hline
\end{tabular*}\vspace*{-3pt}
\end{table}

%
\begin{table}
\tabcolsep=0pt
\caption{Average estimated success
probabilities for top $20$ clusters (ID) in the simulation study, using
full likelihood (lik.) and the split procedure (proc.), percentage of
coverage of the confidence intervals (coverage \%), percentage of times
the true value was less than lower confidence limit [noncov. (above) \%],
and percentage of times the true value was greater than upper
confidence limit [noncoverage (below) \%]}
\label{simprobabilities}
\begin{tabular*}{\textwidth}{@{\extracolsep{4in minus 4in}}ld{2.0}cccccccccccc@{}}
\hline
&&\multicolumn{3}{c}{\textbf{Probability}}&&\multicolumn
{2}{c}{}&&\multicolumn{2}{c}{\textbf{Noncov.}}&&\multicolumn
{2}{c@{}}{\textbf{Noncov.}}\\
&&\multicolumn{3}{c}{\textbf{of success}}&&\multicolumn
{2}{c}{\textbf{Coverage \%}}&&\multicolumn{2}{c}{\textbf{(above) \%}}&&\multicolumn
{2}{c@{}}{\textbf{(below) \%}}\\[-6pt]
&&\multicolumn{3}{c}{\hrulefill}&&\multicolumn
{2}{c}{\hrulefill}&&\multicolumn{2}{c}{\hrulefill}&&\multicolumn
{2}{c@{}}{\hrulefill}\\
\multicolumn{1}{@{}l}{\textbf{Rank}}&\multicolumn{1}{c}{\textbf{ID}}&\textbf{True}&\textbf{Lik.}&\textbf{Proc.}&&\textbf{Lik.}&\textbf{Proc.}
&&\textbf{Lik.}&\textbf{Proc.}&&\textbf{Lik.}&
\multicolumn{1}{c@{}}{\textbf{Proc.}}\\
\hline
\phantom{0}1& 3& 0.72& 0.72& 0.73&& 0.94& 0.95&& 0.02& 0.02&&0.05& 0.04\\
\phantom{0}2& 1& 0.66& 0.66& 0.66&& 0.95& 0.96&& 0.03& 0.02&&0.03& 0.03\\
\phantom{0}3& 33& 0.65& 0.65& 0.65&& 0.98& 0.97&& 0.01& 0.01&&0.02& 0.02\\
\phantom{0}4& 47& 0.64& 0.64& 0.65&& 0.96& 0.96&& 0.02& 0.02&&0.02& 0.02\\
\phantom{0}5& 50& 0.60& 0.60& 0.61&& 0.96& 0.96&& 0.02& 0.02&&0.03& 0.01\\
\phantom{0}6& 27& 0.51& 0.51& 0.51&& 0.96& 0.96&& 0.02& 0.02&&0.03& 0.02\\
\phantom{0}7& 30& 0.51& 0.50& 0.51&& 0.93& 0.94&& 0.03& 0.02&&0.04& 0.03\\
\phantom{0}8& 32& 0.51& 0.50& 0.51&& 0.94& 0.96&& 0.04& 0.02&&0.03& 0.01\\
\phantom{0}9& 14& 0.49& 0.49& 0.49&& 0.97& 0.96&& 0.01& 0.01&&0.03& 0.03\\
10& 7& 0.47& 0.47& 0.47&& 0.94& 0.96&& 0.01& 0.02&&0.05& 0.02\\
11& 9& 0.45& 0.45& 0.45&& 0.97& 0.96&& 0.02& 0.02&&0.02& 0.02\\
12& 48& 0.44& 0.44& 0.44&& 0.96& 0.96&& 0.03& 0.03&&0.01& 0.01\\
13& 10& 0.43& 0.43& 0.43&& 0.92& 0.95&& 0.04& 0.03&&0.05& 0.03\\
14& 21& 0.40& 0.40& 0.40&& 0.97& 0.97&& 0.02& 0.02&&0.01& 0.01\\
15& 11& 0.39& 0.38& 0.39&& 0.95& 0.95&& 0.03& 0.03&&0.03& 0.02\\
16& 26& 0.39& 0.39& 0.39&& 0.94& 0.95&& 0.04& 0.04&&0.02& 0.01\\
17& 15& 0.37& 0.37& 0.37&& 0.96& 0.97&& 0.03& 0.02&&0.01& 0.01\\
18& 13& 0.36& 0.36& 0.36&& 0.95& 0.96&& 0.04& 0.03&&0.02& 0.02\\
19& 4& 0.36& 0.36& 0.36&& 0.94& 0.95&& 0.03& 0.03&&0.04& 0.02\\
20& 42& 0.34& 0.34& 0.34&& 0.95& 0.97&& 0.04& 0.02&&0.02& 0.01\\
\hline
\end{tabular*}\vspace*{-3pt}
\end{table}

Further scrutiny of the estimated success probabilities in Table~\ref{simprobabilities} confirms the similarity in performance between the
two methods. Here again the point estimates are very close to the true
values, and the coverage of the confidence intervals lies around 95\%\vadjust{\goodbreak}
for maximum likelihood as well as for the split procedure. Relative
differences between the true values and estimates from the two methods
are mostly positive, suggesting that many cluster effects were slightly
overestimated. These results further confirm the heuristic conclusions
derived in Section~\ref{sec:method}, stating that the split procedure
should often yield results very similar to maximum likelihood when $W$
is sufficiently large.\vadjust{\goodbreak}

\section{Conclusion}\label{sec:conclusion}
In our quest to quantify expert opinion on the potential of clusters
of chemical compounds, we have introduced a \emph
{permutational-splitting sample procedure}. A combination of maximum
likelihood estimation, resampling and stochastic methods produced
parameter estimates and confidence intervals comparable to those
obtained from full maximum likelihood. Loss in precision with the split
procedure, apparent in wider confidence intervals, is anticipated,
because the procedure splits the data into dependent subsamples,
resulting in a less efficient estimate of the random-effect variance.\vadjust{\goodbreak}

The model used for the statistical analysis and the conclusions derived
from it rest on a number of assumptions, such as the distribution of
the expert-specific effect $b_i$. Although the normality assumption for
the random effects is standard in most software packages, in principle,
it would be possible to consider other distributions. For instance,
using probability integral transformations in the SAS procedure
NLMIXED, other distribution could be fitted, but obtaining convergence
is much more challenging with these models [\citet{Nelson}].

One could also extend the model by letting the expert effects vary
among clusters. However, this extension would dramatically increase the
dimension of the vector of random effects, aggravating the already
challenging numerical problems. In general, the successful application
of the Rasch model in psychometrics to tackle problems similar to the
one considered here makes us believe that, although it cannot be
formally proven, model (\ref{eq:modelmain}) may offer a feasible and
reliable way to estimate the success probabilities of interest.

More simulation studies and applications to real problems will shed
light on the potential and limitations of the model and fitting
procedure proposed in the present work. Importantly, their application
is possible with commonly available software, and a simulated data set
with the corresponding SAS code for the analysis can be freely
downloaded from \url{http://www.ibiostat.be/software/}.

Even though it was not the focus of the present work, it is clear that
the design of the study is another important element to guarantee the
validity of the results. Optimal designs are a class of experimental
designs that are optimal with respect to some statistical criterion
[\citet{Berger2009}]. For instance, one may aim to select the number of
experts, the number of clusters assigned to the experts and the
assignment mechanism to maximize precision when estimating the
probabilities of success. In principle, it seems intuitively desirable
for each cluster to be evaluated by the same number of experts and for
each pair of experts to have a reasonable number of clusters in common.
However, more research will be needed to clarify these issues and
establish the best possible design for this type of study.

\section*{Acknowledgments}We
kindly acknowledge the following colleagues at Johnson \& Johnson for
generating and providing the data set: Dimitris Agrafiotis, Michael
Hack, Todd Jones, Dmitrii Rassokhin, Taraneh Mirzadegan, Mark
Seierstad, Andrew Skalkin, Peter ten Holte and the Johnson \& Johnson
chemistry community.

The authors are deeply grateful to the Associate Editor for offering
outstanding advice and suggestions, which have led to a major
improvement of the manuscript.

For the computations, simulations and data processing, we used the
infrastructure of the VSC Flemish Supercomputer Center, funded by the
Hercules Foundation and the Flemish Government department EWI.




\printaddresses
\end{document}